# *Mineralogy, chemistry and composition of organic compounds in the fresh carbonaceous chondrite Mukundpura: CM1 or CM2?*

S. Potin[1], P. Beck[1], L. Bonal[1], B. Schmitt[1], A. Garenne[2], F. Moynier[3], A. Agranier[4], P. Schmitt-Kopplin[5,6], A.K. Malik[7], E. Quirico[1]

[1]Université Grenoble Alpes, CNRS, Institut de Planétologie et d'Astrophysique de Grenoble IPAG, France (414 rue de la Piscine, 38400 Saint-Martin d'Hères, France,

[2]Nuclear and Chemical Sciences Division, Lawrence Livermore National Laboratory, 7000 East Avenue, Livermore, CA 94550, USA

[3]Institut de Physique du Globe de Paris IPGP (1 Rue Jussieu, 75005 Paris, France)

[4]Laboratoire Géosciences Océan UMR/CNRS 6538, IUEM, Université de Bretagne Occidentale (Technopôle Brest-Iroise, Rue Dumont d'Urville, 29280 Plouzané, France)

[5]Helmholtz Zentrum Muenchen, Research Unit Environmental Simulation (EUS) Ingolstaedter Landstrasse 1, 85764 Neuherberg

[6]Technische Universität München, Lehrstuhl für Analytische Lebensmittechemie, Maximus-von-Imhof-Forum 2, 85354 Freising, Germany

[7]Punjabi University, Department of Chemistry, Patiala, Patiala - 147 002, Punjab, India

ABSTRACT

We present here several laboratory analyses performed on the freshly fallen Mukundpura CM chondrite. Results of infrared transmission spectroscopy, thermogravimetry analysis and reflectance spectroscopy show that Mukundpura is mainly composed of phyllosilicates. The rare earth trace elements composition and ultrahigh resolution mass spectrometry of the soluble organic matter (SOM) give results consistent with CM chondrites. Finally, Raman spectroscopy shows no signs of thermal alteration of the meteorite. All the results agree that Mukundpura has been strongly altered by water on its parent body. Comparison of the results obtained on the meteorite with those of other chondrites of known petrologic types lead to the conclusion that Mukundpura is similar to CM1 chondrites, which differs from its original classification as a CM2.

1. Introduction

Meteorite provide a natural sampling of Solar System small bodies, and as a corollary our understanding of the small bodies population strongly relies on what type of samples are available for laboratory studies. Spectroscopic observations have revealed that roughly half of the asteroids main-belt is composed of dark objects belonging to the C-complex (DeMeo and Carry 2013). However, when looking at the available suite of extra-terrestrial materials, carbonaceous chondrites are rare among falls (4 % of chondrites, (Krot et al. 2006)) and even rarer among finds (3 % of chondrites, (Krot et al. 2006)). Meteorites belonging to the CM families are of particular interest to the community, because these samples preserve extensive evidence for fluid-rock interactions in the form of secondary minerals including phyllosilicates, carbonates and sulfides. They record an early hydrothermal process, they preserve some of the "water" that was accreted on their parent body in the form of $OH/H_2O$ in minerals, and they are particularly rich in organic compounds when compared to other chondrites classes. The putative parent bodies of CM chondrites are Ch and Cgh type asteroids, and there are two ongoing sample return mission at the time of the writing targeting possible parent bodies of CM chondrites (Hamilton et al. 2019; Kitazato et al. 2019). There are at present 17 CM chondrites fall in the meteorite bulletin database with 13 of them from the 19 and

20 the centuries. Because terrestrial residence may impact mineralogy and organo-chemistry of extra-terrestrial samples, the rare possibility to investigate fresh CM chondrites has motivated this study.

The Mukundpura meteorite fell in India on June 6, 2017 (Tripathi et al. 2018) and was immediately collected by the authorities, reducing the post-fall terrestrial alteration. Initial petrographic observations of the newly fallen meteorite were performed: the meteorite is mostly composed of $Fe_2O_3$ (32.4 wt%), $SiO_2$ (29.5 wt%) and MgO (17.4 wt%), and contains in volume 70% of matrix, 15% of chondrules and 2% of refractory inclusions (Ray and Shukla 2018). The matrix/inclusion ratio and the petrographic analysis were found to be consistent with a CM chondrite (Kallemeyn and Wasson 1981; Van Schmus and Wood 1967). Mukundpura was first classified as a CM2 chondrite (Ray and Shukla 2018), then as a CM1 chondrite (Rudraswami et al. 2019). The attribution of petrologic types 1 and 2 were both established based on the bulk chemical composition and petrographic analyses.

The freshness of the Mukundpura fall allows investigations of its post-accretion aqueous and thermal history with minimal interfering tertiary products from terrestrial weathering. The mineralogy, trace elements, optical properties and organic composition of the meteorite were analyzed in the present work and compared to results obtained on other chondrites of different classifications and alteration histories. Our results will be used to discuss similarities and differences between Mukundpura and other CM chondrites in term of alteration mineralogy and bulk composition. We will also try to decipher the thermal history of the sample using Raman spectroscopy on macromolecular organics. Last, we will assess whether organics compounds in a fresh meteorite fall are different from those present in other samples of CM chondrites.

2. Samples and analytical methods

   a. Petrology based on studies of Mukundpura

There are three previous publications that looked at the petrography of Mukundpura (Ray and Shukla 2018; Rudraswami et al. 2019; Tripathi et al. 2018). All this studies concluded that the sample is a CM chondrite but the reported extents of aqueous alteration differ. Ray and Shukla (2018) concluded that the sample was only modestly altered (CM2), while Rudraswami et al. (2019) concluded that Mukundpura is at the boundary between CM2 and CM1. All studies confirm that the fine-grained matrix of the samples consist of abundant phyllosilicates, poorly cristallized phases (PCP) and carbonates (calcite mostly). Chondrules are present but the areal fraction of chondrules varies between 15 % (Ray and Shukla 2018) and 7 % (Rudraswami et al. 2019). In the case of Rudraswami et al. (2019) while the fraction of chondrules identified was low (lower that typical CM chondrites), a number of relict chondrules were identified. This difference in the petrography of the samples can be related to the brecciated nature of CM chondrites and could explain the difference classification reached by different authors.

   b. Samples

Two samples of Mukundpura were considered here: the first sample (Figure 1a, called sample A) was mainly used for the characterization of the alteration history and bidirectional reflectance spectroscopy, and the second (Figure 1b called sample B) was used for the mass spectrometry. The two samples were raw broken pieces of the meteorite. The samples A and B weighed approximately 200mg and 500mg respectively.

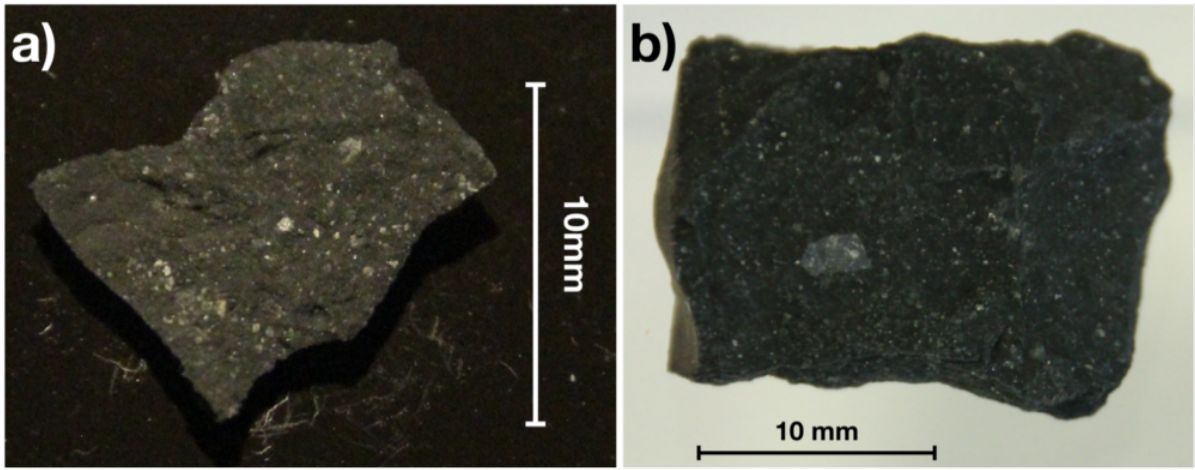

Figure 1: Pictures of the two samples of Mukundpura. a) Bulk sample characterized through each of the analytical methods, b) Sample characterized through high-resolution mass spectrometry.

The sample A was manually dry crushed in an agate mortar. The resulting powder was used for the thermo-gravimetric analysis (TGA), Raman, Mid-infrared (mid-IR) transmission and visible-near infrared (NIR) reflectance spectroscopy.

c. Thermo-gravimetric analysis

Thermo-gravimetric analysis (TGA) and differential scanning calorimetry DSC (TGA-DSC3+ Mettler-Toledo) were performed at the Institut des Sciences de la Terre (ISTerre, Grenoble - France) following the protocol described in Garenne et al. (2014). Data were acquired from room temperature to 1100°C, with a heating rate of 10°C/min and a mass resolution of 1µg over the whole range, under a continuous flux of gaseous inert nitrogen to avoid the reaction of the sample with oxygen during heating. An initial mass of 15.263mg was taken from the powdered sample A to perform the analysis.

d. Mid-IR transmission spectroscopy

About 20 mg were taken from the crushed sample (Sample A). Out of this mass, 1.0 mg was weighed and mixed with 300 mg of commercial ultrapure potassium bromide powder (KBr). This mixture was then compressed to 400 bars in order to obtain a 13 mm diameter pellet. Mid-IR spectra were measured with a Brucker V70v spectrometer at the Institut de Planétologie et d'Astrophysique de Grenoble (IPAG, Grenoble - France), following the method described in previous studies (Beck et al. 2014). Spectra were acquired with a spectral resolution of 2 $cm^{-1}$ in the 2000-400 $cm^{-1}$ range (5 to 25 µm). The shape of the 3.0µm water band of the obtained spectra is not only controlled by the hydration state of the sample but also by the water contained in the hydrophilic KBr. Thus, it will not be interpreted here.

e. Visible-Near IR Reflectance spectroscopy

Reflectance spectroscopy of sample A was conducted at IPAG with the spectro-gonio radiometer SHADOWS (Potin et al. 2018). As the sample was ground and not sieved, the powder contained a large distribution of grain sizes but no larger than a few hundreds of microns. Then, the spectrum was acquired from 400 nm to 4000 nm under a nadir illumination (incidence 0° normal to the surface of the sample) and an observation angle of 30° from the normal of the surface. The spectral resolution of the instrument during the reflectance measurement varies with the spectral range and is given in Table 1.

| Spectral range (nm) | Spectral resolution (nm) |
| --- | --- |
| 400- 679 | 4.85 – 4.75 |
| 680 – 1400 | 9.71 – 9.38 |
| 1500– 2999 | 19.42 – 18.73 |

| 3000– 4000 | 38.84 – 38.44 |

Table 1: Spectral resolution of the reflectance spectra.

The bright xenolith present in sample B was large enough to be studied using the goniometer's reduced illumination spot (Potin et al. 2018). The spectral resolution of the instrument is the same as displayed in Table 1.

f. Raman spectroscopy

Raman spectra were performed on matrix grains from the sample A of Mukundpura. Grains were collected from the matrix under an optical binocular based on texture and color with polarized light, and pressed between two glass slides, similarly to the method described in Bonal et al. (2016). The measured Raman spectra showed spectral bands related to the presence of polyaromatic carbonaceous matter, which confirms *a posteriori* the successful selection of matrix grains. Spectra were acquired using a Labram spectrometer (Horiba-Jobin-Yvon) at the Laboratoire de Géologie de Lyon (ENS Lyon - France). This instrument is equipped with a Spectra Physics Argon ion laser excited at 514.5 nm. The laser beam passes through a 100x objective, leading to a spot of approximately 2 µm in diameter. The experimental conditions and analytical methods were similar to the ones used by Quirico et al. (2018) to characterize type 1 and 2 chondrites. The laser power on the surface sample was 300 ± 20 µW, as measured out of a 10x objective. A total of 35 spectra on 13 individual matrix grains were acquired in the spectral range between 500 cm$^{-1}$ and 2230 cm$^{-1}$.

g. Trace-element analysis

100 mg of sample was crushed in an agate mortar, formerly cleaned in 4N $HNO_3$. Subsequently, two aliquots of powder about 20mg each (fractions 1 and 2) were dissolved in 2mL of 24N distilled HF + 0.5 mL 12N suprapur $HClO_4$ within polytetrafluoroethene (PTFE) vials, placed inside a spring loaded screw-top steel vessel (Parr bombs, see Navarro et al. (2008) for details) at 150°C for 7 days. After evaporation, each sample was re-dissolved in 1 mL of 14N distilled $HNO_3$ and split in three distinct fractions, each containing ~6.7 g of sample (Mukundpura 1-1 to Mukundpura 1-3 and Mukundpura 2-1 to Mukundpura 2-3). After a second evaporation, the 6 fractions and three basaltic certified georeference standards (BIR-1, BCR2 and $BHVO_2$) digested in a similar manner, were re-dissolved in 1g of a 0.5N $HNO_3$ mixture containing 1ppb of In (Li and Lee 2006).

Measurements were obtained using the High Resolution Inductively-Coupled Plasma Mass Spectrometer (HR-ICP-MS) Element XR (ThermoFisher Scientific) at Institut Universitaire Européen de la Mer (Brest University) connected to a nitrogen supplied desolvating Nebulizer (apex Q Elemental Scientific) introduction system.

All samples were externally calibrated using calibration standard solutions (0.1, 0.2, 0.3 and 0.7ppb) and the accuracy and external reproducibility of the data were monitored by repeated measurements of the georeference standards BCR2, BHVO2 and BIR-1.

h. Ultrahigh resolution mass spectrometry of the soluble organic matter

The soluble organic compounds extracted from Sample B were characterized by electrospray ionization ultrahigh resolution mass spectrometry (12 Tesla ion-cyclotron-resonance Fourier-transform mass-spectrometry, FTICR-MS) at the HelmholtzZentrum muenchen (German Research Center for Environmental Health). The sample-conditioning and the procedure of electrospray ionization analysis are similar to the ones described for the CM2 Murchison in Schmitt-Kopplin et al. (2010).

3. Results on the aqueous alteration experienced by Mukundpura
    a. Thermo-gravimetric analysis

TGA is used to quantify the amount of water and –OH groups. The sample is gradually heated to 1100°C, and variation of its mass are monitored through the experiment. Each mass variation is typical from a specific type of water or –OH group, thus the experiment can be separated into different temperature ranges (Garenne et al. 2014): (1) from room temperature to 200°C, the mass loss corresponds to the release of adsorbed molecular water and mesopore water; (2) from 200°C to 400°C, the mass loss is mostly controlled by the release of water contained in oxy-hydroxide minerals such as ferrihydrite $Fe_2O_3 \bullet 0.5H_2O$; (3) from 400°C to 770°C, the mass loss corresponds to the release of -OH groups from phyllosilicates. Note that some overlap exists between decomposition Fe-rich hydroxide and some phyllosilicates (King et al. 2015). Still this segmentation in temperature ranges provides a convenient way to compare various meteorite samples.

Figure 2 presents the TGA results, showing the evolution of the mass of the Mukundpura sample with the increasing temperature. The derivative of the curve is also plotted as it allows a precise determination of the temperature at which the mass loss occurs and so the identification of the host hydrated mineral.

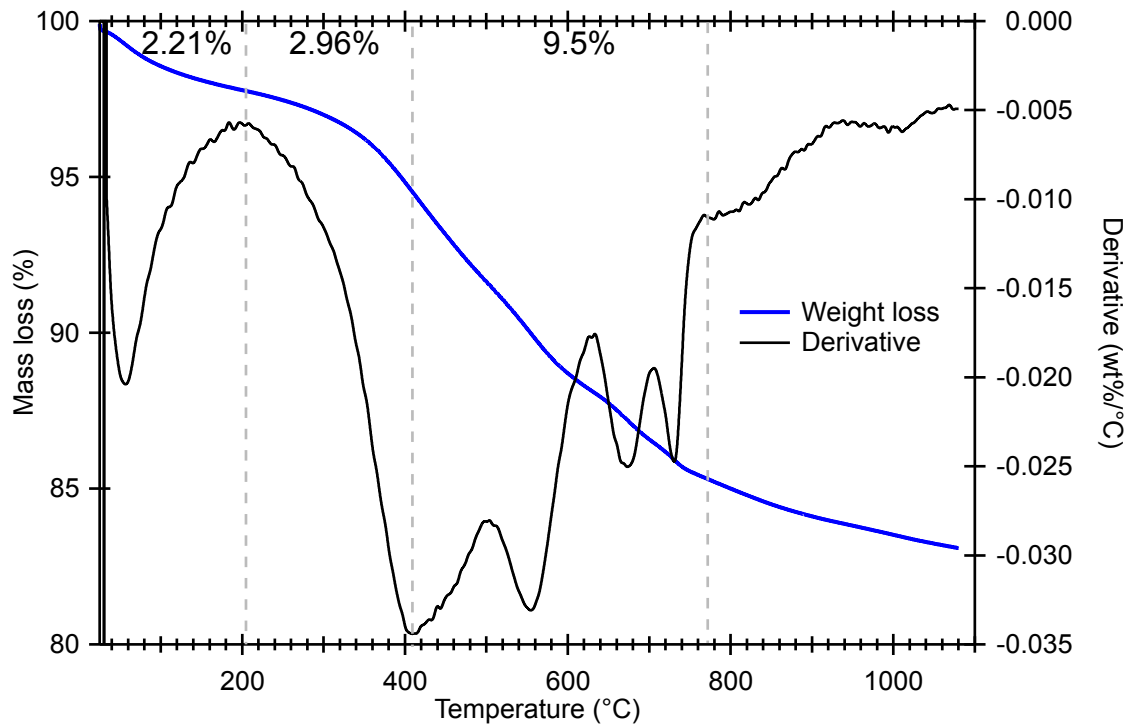

Figure 2: Thermogravimetric analysis of Mukundpura, showing the mass loss (broad blue curve) and the derivative (thin black curve). The grey doted lines separate the temperature ranges described above. Mass losses for the first 3 ranges are also noted.

Between room temperature and 200°C, the mass loss is of 2.21 wt%. This thermal range corresponds to the release of both the absorbed molecular water, around 80°C, and the mesopore water above 100°C. The derivative line shows only one peak at 80°C, corresponding to absorbed molecular water. No trace of mesopore water is seen in Figure 2. Roughly 3 wt. % of the water is contained in the 200C-400C range. This range is mainly associated to dehydration oxyhydroxides minerals. However, we think this value can be overestimated as fully associated to hydroxides and small contribution of the main minerals (e.g. phyllosilcates) is also expected.

Note that Tripathi et al. (2018) observed very similar mass loss, 2.0wt% from 200-400°C, and 9.8wt% from 400-770°C.

b. Transmission IR spectroscopy

The degree of aqueous alteration can be assessed by analysing the shape of the $SiO_4$ stretching band around 10 µm, and the detection/absence of olivine signatures at 11.2 µm and 19.5 µm (Beck et al. 2014). Aqueously altered CM chondrites present the phyllosilicate band at 10 µm with faint or undetectable olivine signatures, and inversely unaltered chondrites present the olivine signatures without the phyllosilicate bands. The transmission IR spectrum of bulk Mukundpura is presented in Figure .

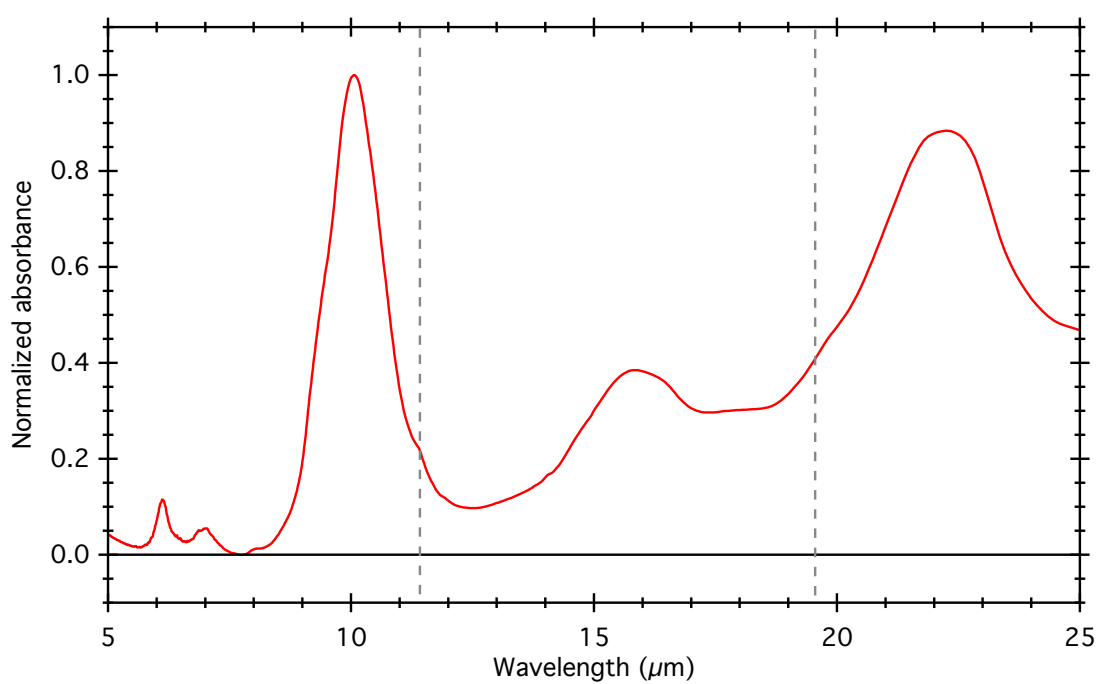

Figure 3: Mid-infrared transmission spectrum of the bulk of Mukundpura. The dotted grey lines indicate the position of the olivine signatures at 11.2 and 19.5µm. The spectrum was normalized to the intensity of the 10µm band.

The strong $SiO_4$ stretching band at 10 µm indicates the presence of phyllosilicates, and faint olivine signatures are still detected at 11.2 and 19.5µm. This is in favour of a intensive water alteration on the parent body, where the majority of the initial mafic silicates have been transformed into phyllosilicates (Beck et al. 2014). Note that the Mukundpura spectrum does not show strong evidence of the presence of carbonates at 7μm as can be observed for instance in CM chondrites (Knacke and Krätschmer 1980).

4. Results on the thermal alteration history of Mukundpura

Figure 4 presents the raw Raman spectra obtained on matrix grains of Mukundpura.

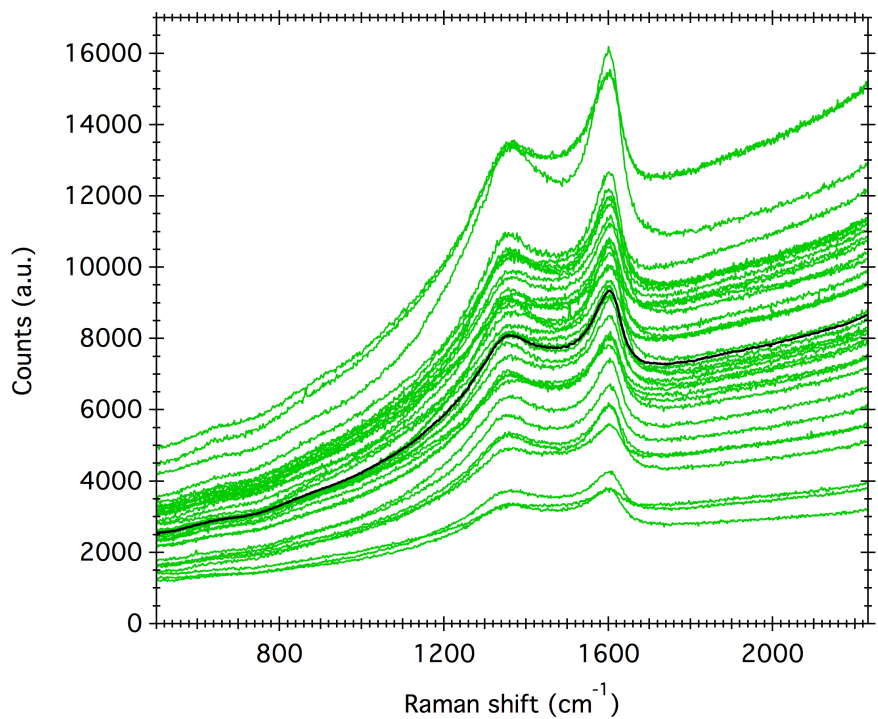

Figure 4: Raman spectra of the matrix of the Mukundpura meteorite (green) and average spectrum (black).

All spectra show the so-called D- and G-bands around 1350 and 1600 $cm^{-1}$ related to the presence of polyaromatic carbonaceous matter in the matrix of Mukundpura. A significative slope due to the background fluorescence is superimposed upon the Raman signatures. This effects is characteristic of unheated or weakly thermally altered chondrites (Quirico et al. 2014). To extract Raman peak parameters, the fluorescence background has been removed from the data using a first order linear fit between 800cm$^{-1}$ and 2000 cm$^{-1}$. Then the D- and G-band are fitted by a Lorentzian profile and a Breit-Wigner-Fano (BWF) profile, respectively (e.g. Bonal et al. 2016). The peak position, intensity and full width at half maximum are calculated for each of the bands on all spectra normalized to the G-band. The average value and the standard deviation (1σ) are displayed in Table 2.

|  | FWHM$_G$ (cm$^{-1}$) | w$_G$(cm$^{-1}$) | I$_D$/I$_G$ | FWHM$_D$ (cm$^{-1}$) | w$_D$ (cm$^{-1}$) |
|---|---|---|---|---|---|
| Mukundpura | 102.6 ± 7.3 | 1594.3 ± 1.1 | 1.041± 0.087 | 245.1 ± 16.0 | 1360.6 ± 2.5 |
| Unheated CM (range) | 90 - 115 | 1586 - 1596 | 0.85 – 1.0 | 210 - 260 | 1357 - 1366 |
| Heated CM (range) | 70-110 | 1587-1605 | 0.7-1.3 | 160-205 | 1357-1372 |

Table 2: Raman spectral parameters of the Mukundpura meteorite and parameters range for unheated CM chondrites (data from Quirico et al. (2014)).

Raman spectral parameters of the G-band show that the band is centered around 1594 cm$^{-1}$ and is rather broad with a FWHM reaching 102.6cm$^{-1}$. Compared to the ranges described in Quirico et al. (2014) for unheated CM chondrites, the Raman parameters of Mukundpura indicate that the meteorite is a primitive chondrite having escaped major heating or shock-related metamorphism.

5. Trace-elements composition

The results of the trace element analysis are presented in Table 3 and illustrated in Figure 5.

| | Mukundpura fraction 1 | | | Mukundpura fraction 2 | | | Geostandards | | |
|---|---|---|---|---|---|---|---|---|---|
| | | | | | | | BHVO2 (average, n=3) | BIR1 (average, n=3) | BCR-2 (average, n=3) |
| HR-ICP-MS Low Resolution (ppm) | | | | | | | | | |
| Li | 1.36 | 1.36 | 1.37 | 1.47 | 1.51 | 1.48 | 4.64 | 2.82 | 9.55 |
| Be | | | | | | | 1.13 | 0.121 | 2.21 |
| Sc | 20.3 | 25.7 | 28 | 19.1 | 26.7 | 27.6 | 68.7 | 50.7 | 11.5 |
| Ti | 451 | 455 | 455 | 435 | 434 | 431 | 15640 | 5370 | 12680 |
| V | 56.9 | 58.2 | 57.5 | 56 | 56.1 | 54.8 | 302 | 301 | 384 |
| Cr | 2209 | 2227 | 2202 | 2104 | 2142 | 2100 | 147 | 229 | |
| Co | 410 | 413 | 419 | 410 | 413 | 408 | 39.3 | 45.9 | 26.1 |
| Ni | 11470 | 10970 | 11400 | 10770 | 10800 | 10870 | 119 | 170 | 13.3 |
| Cu | 87.5 | 87 | 87.4 | 84.8 | 85.4 | 84.8 | 128 | 117 | 24 |
| Zn | 224 | 219 | 222 | 222 | 228 | 228 | 139 | 37.4 | 98.8 |
| Ga | 6.2 | 6.29 | 6.32 | 6.24 | 6.29 | 6.25 | 22.9 | 15.2 | 22.9 |
| Ge | 4.76 | 5.07 | 5.44 | 5.83 | 6.19 | 6.52 | 3.45 | 2.46 | 3.76 |
| Rb | 1.44 | 1.46 | 1.46 | 1.45 | 1.48 | 1.49 | 9.48 | 0.371 | 47.5 |
| Sr | 8.46 | 8.44 | 8.49 | 8.23 | 8.4 | 8.26 | 410 | 112 | 348 |
| Y | 1.53 | 1.55 | 1.54 | 1.37 | 1.37 | 1.35 | 27.5 | 16.6 | 37.4 |
| Zr | 4.74 | 4.37 | 4.51 | 4.28 | 4.31 | 4.27 | 169 | 14.6 | 182 |
| Nb | 0.405 | 0.378 | 0.385 | 0.353 | 0.366 | 0.357 | 18.6 | 0.569 | 12.3 |
| Cs | 0.103 | 0.1 | 0.103 | 0.0988 | 0.1 | 0.101 | 0.0979 | 0.0018 | 1.13 |
| Ba | 2.54 | 2.54 | 2.53 | 2.98 | 3.04 | 2.98 | 135 | 6.76 | 694 |
| La | 0.254 | 0.249 | 0.245 | 0.234 | 0.238 | 0.235 | 13.6 | 0.543 | 22.3 |
| Ce | 0.651 | 0.652 | 0.641 | 0.625 | 0.651 | 0.621 | 35 | 1.77 | 49.2 |
| Pr | 0.0956 | 0.0984 | 0.0971 | 0.097 | 0.0995 | 0.0973 | 5.38 | 0.376 | 6.85 |
| Nd | 0.484 | 0.49 | 0.493 | 0.485 | 0.5 | 0.484 | 25.5 | 2.46 | 30 |
| Sm | 0.157 | 0.156 | 0.162 | 0.153 | 0.163 | 0.154 | 6.39 | 1.14 | 6.92 |
| Eu | 0.0587 | 0.0594 | 0.0587 | 0.0575 | 0.0593 | 0.0589 | 2.17 | 0.537 | 2.12 |
| Eu | 0.0596 | 0.0587 | 0.0599 | 0.0575 | 0.0602 | 0.0587 | 2.15 | 0.528 | 2.18 |
| Tb | 0.0368 | 0.0369 | 0.0375 | 0.0365 | 0.037 | 0.0364 | 1.01 | 0.371 | 1.14 |
| Gd | 0.21 | 0.213 | 0.214 | 0.202 | 0.207 | 0.206 | 6.63 | 1.98 | 7.26 |
| Dy | 0.247 | 0.243 | 0.244 | 0.24 | 0.242 | 0.248 | 5.68 | 2.64 | 6.8 |
| Ho | 0.0542 | 0.0541 | 0.0537 | 0.0509 | 0.0524 | 0.0539 | 1.04 | 0.598 | 1.35 |
| Er | 0.163 | 0.161 | 0.154 | 0.157 | 0.152 | 0.154 | 2.66 | 1.76 | 3.82 |
| Tm | 0.0247 | 0.0251 | 0.0261 | 0.0236 | 0.0246 | 0.0243 | 0.351 | 0.265 | 0.555 |
| Yb | 0.177 | 0.168 | 0.172 | 0.162 | 0.163 | 0.162 | 2.16 | 1.78 | 3.58 |
| Lu | 0.0242 | 0.0235 | 0.0237 | 0.0226 | 0.023 | 0.0238 | 0.293 | 0.258 | 0.517 |
| Hf | 0.111 | 0.105 | 0.108 | 0.106 | 0.105 | 0.108 | 4.42 | 0.582 | 4.87 |
| Ta | 0.0911 | 0.0697 | 0.0689 | 0.0499 | 0.061 | 0.0556 | 1.27 | 0.0514 | 0.855 |
| Tl | 0.0685 | 0.0664 | 0.0676 | 0.0632 | 0.066 | 0.0643 | 0.0226 | | 0.3 |
| Pb | 1.3 | 1.3 | 1.32 | 1.3 | 1.34 | 1.33 | 1.55 | 3.58 | 11.8 |
| Th | 0.0337 | 0.032 | 0.0311 | 0.0278 | 0.0287 | 0.0281 | 1.04 | 0.0223 | 5.06 |
| U | 0.0116 | 0.01 | 0.0104 | 0.0095 | 0.0109 | 0.0106 | 0.422 | 0.0131 | 1.73 |
| Cd | 0.227 | 0.225 | 0.227 | 0.237 | 0.238 | 0.238 | 0.132 | 0.0859 | 0.515 |
| Ag | 0.0131 | 0.0128 | 0.0129 | 0.0118 | 0.0125 | 0.012 | 0.113 | 0.0031 | 0.113 |
| Pt | 0.129 | 0.129 | 0.128 | 0.127 | 0.129 | 0.127 | 0.0109 | 0.002 | 0.0128 |
| Au | 0.0042 | 0.0038 | 0.0037 | 0.0136 | 0.0136 | 0.0136 | 0.0014 | | |
| Sn | 0.0055 | 0.0055 | 0.0057 | 0.0063 | 0.0064 | 0.0064 | 0.0151 | 0.0054 | 0.0169 |
| Sb | 0.085 | 0.0778 | 0.0749 | 0.0716 | 0.0745 | 0.0704 | 0.0721 | 0.425 | 0.259 |
| HR-ICP-MS Medium Resolution (ppm) | | | | | | | | | |
| Na | 3545 | 3629 | 3637 | 3606 | 3577 | 3589 | 16460 | 13530 | 22990 |
| Mg | 92610 | 93620 | 94020 | 90360 | 89730 | 89410 | 47800 | 62620 | 23140 |
| Al | 7177 | 7435 | 7471 | 5563 | 5508 | 5460 | 71760 | 82260 | 61700 |
| P | 380 | 355 | 324 | 291 | 283 | 281 | 432 | 26.7 | 578 |
| K | 394 | 411 | 417 | 330 | 328 | 327 | 4253 | 189 | 15160 |
| Ca | 9099 | 9222 | 9256 | 8982 | 8837 | 8812 | 80590 | 93720 | 48640 |
| Sc | 5.68 | 5.73 | 5.71 | 4.36 | 4.29 | 4.22 | 31.2 | 42.6 | 32 |
| Ti | 449 | 452 | 448 | 431 | 426 | 425 | 15690 | 5444 | 12560 |
| V | 51.4 | 52 | 51.6 | 48.9 | 48.6 | 48.1 | 302 | 302 | 389 |
| Cr | 2175 | 2211 | 2209 | 2121 | 2102 | 2110 | 275 | 379 | 13.3 |
| Mn | 1257 | 1274 | 1272 | 1243 | 1235 | 1240 | 1325 | 1336 | 1517 |
| Fe | 152600 | 154500 | 154900 | 153300 | 152800 | 153800 | 86880 | 78950 | 95000 |
| Co | 408 | 414 | 417 | 411 | 407 | 408 | 45.5 | 52.9 | 37.6 |
| Ni | 8031 | 8116 | 8124 | 7975 | 7922 | 7981 | 116 | 165 | 11 |
| Cu | 101 | 101 | 101 | 99.1 | 98.8 | 98 | 139 | 133 | 18.1 |
| Zn | 125 | 126 | 128 | 128 | 128 | 129 | 106 | 75.2 | 143 |
| Zr | 3.99 | 3.88 | 3.98 | 3.81 | 3.71 | 3.73 | 157 | 13.3 | 168 |
| Nb | 0.328 | 0.32 | 0.322 | 0.295 | 0.305 | 0.3 | 16.2 | 0.501 | 11.1 |
| U | 0.0109 | 0.0106 | 0.0087 | 0.0103 | 0.0102 | 0.0102 | 0.433 | 0.0143 | 1.73 |
| Cd | 0.207 | 0.233 | 0.216 | 0.229 | 0.214 | 0.232 | | | |
| Ag | 0.0139 | 0.0137 | 0.0136 | 0.0132 | 0.0123 | 0.0121 | 0.117 | 0.0079 | 0.121 |

Table 3: Trace element composition of the Mukundpura meteorite.

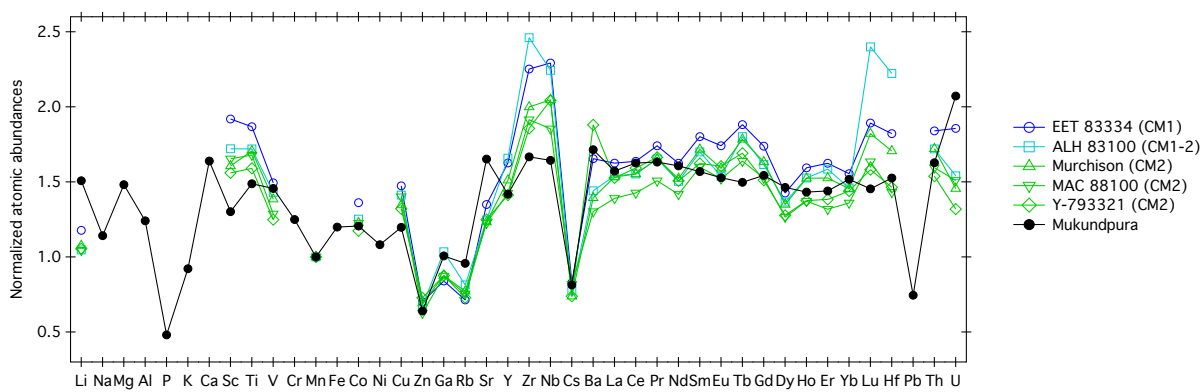

Figure 5: Atomic abundances of Mukundpura compared to other aqueously altered CM chondrites (data from Friedrich et al. (2002)). The values have been normalized to Orgueil (data from Barrat et al. (2012)), then Mn-normalized.

Mukundpura is slightly enriched in rare earth elements (REE) (1.2) when compared to CI, with a flat REE pattern. The Zn/Mn vs Sc/Mn array has been proved useful to distinguish samples belonging to different carbonaceous chondrite groups (Boynton 1984). This array shows that Mukundpura has a Zn/Mn ratio in agreement with CM chondrites and a Sc/Mn ratio slightly lower but still in agreement with values obtained for CM chondrites. The standards demonstrated external reproducibilities better than 5–10% depending on the element and concentration.

6. Composition of the solvent soluble organic compounds

The analysis of the methanol soluble organic fraction revealed a high chemical diversity of the C, H, O, N, S bearing compounds in a similar polarity range compared to the Murchison CM2; the obtained van Krevelen diagrams visualize Mukundpura compared to Murchison in the H/C vs O/C and the H/C vs. m/z spaces, with similar compositional profiles and almost identical proportions in the abundance in CHO, CHNO,

CHOS and CHNOS molecules (Figure 6). The only difference is seen in the lower mass range covered by Mukundpura resulting in an lower absolute number in mass signals and counted formula (7.500 vs. 13.000 for Murchison). These two CM were compared in Figure 6 to WIS 91600, as a classical thermally altered meteorite presented already with TGA in Table 4. High thermal alteration fully destroyed the profiles of the soluble and only few hundred compound (1.700) remaining are within the CHO and CHOS compounds. This was shown earlier with the fresh fall Sutter´s Mill (Jenniskens et al. 2012) and confirms that Mukundpura was not affected by high temperatures.

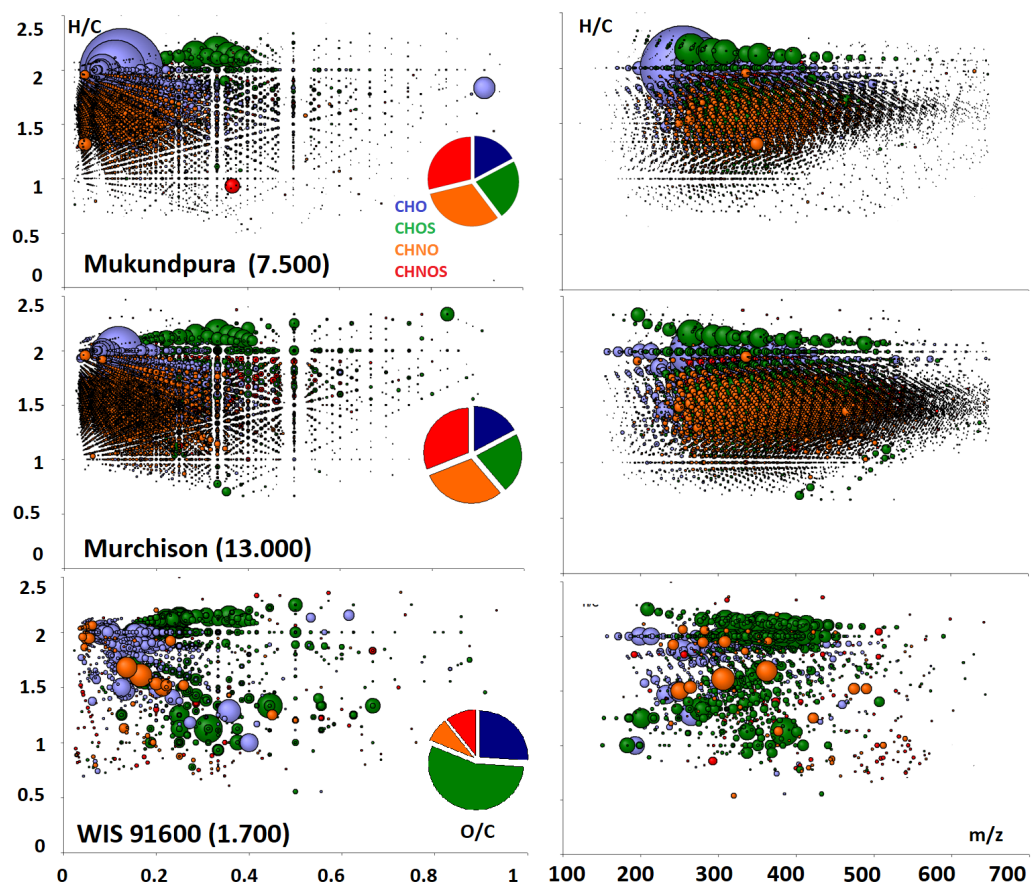

Figure 6: van Krevelen diagrams (H/C vs. OC) and (H/C vs. m/z) calculated from the FTICR-mass spectrometric ultrahigh resolution mass profiles of Mukundpura compared to Murchison CM2 and WIS 96100 thermaly altered CM2. Each individual bubble

corresponds to a formula calculated from exact masses with a size proportional to its abundance and a colour relative to the presence of elements.

7. Optical spectroscopy
    a. Reflectance spectroscopy of the bulk sample

The reflectance spectrum of the Mukundpura meteorite is compared to spectra from other aqueously altered chondrites in Figure 7. The spectrogonio-radiometer SHADOWS was used as well to acquire the reflectance spectra of the other chondrites, using the same geometrical configuration and spectral resolution (Potin et al., under review).

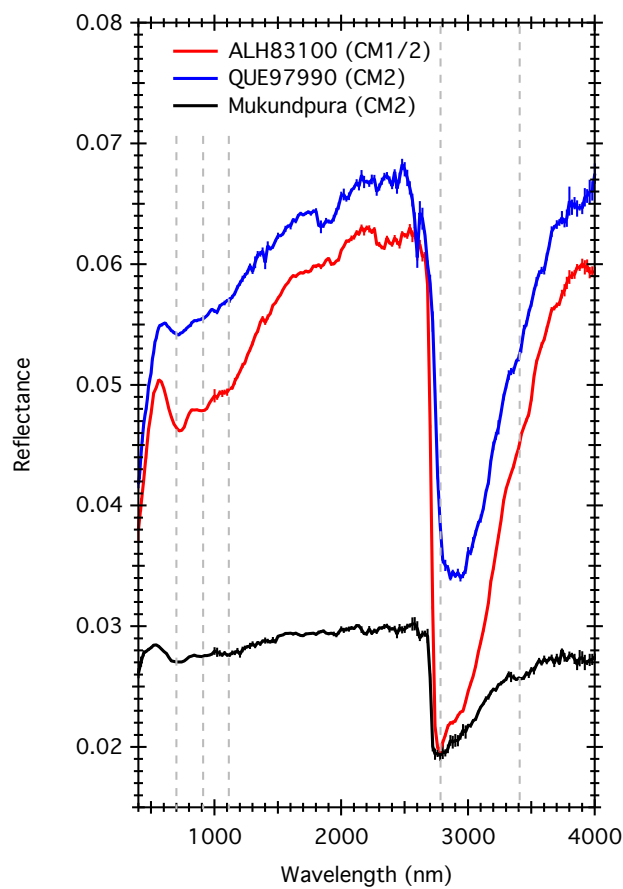

Figure 7: Reflectance spectroscopy of the Mukundpura meteorite compared to other aqueously altered chondrites. The errorbars are displayed on each spectrum. The grey

lines mark the position of the typical CM2 absorption bands at 700nm, 900nm, 1100nm, 2780nm and 3400nm.

The reflectance spectrum of Mukundpura shows typical spectral characteristics of aqueously altered CM chondrites (Cloutis et al. 2011): (1) a rather low general reflectance around 0.028, (2) an absorption feature around 700nm corresponding to the $Fe^{2+}$-$Fe^{3+}$ charge transfer in Fe-bearing phyllosilicates, (3) features at 900 and 1100nm due to $Fe^{2+}$ crystal field transisions, (4) a faint but detectable organic feature around 3400nm and (5) a strong absorption band around 2750nm due to stretching –OH hydroxyls in Mg-bearing phyllosilicates. The –OH band usually occurs around 3000nm and its position depends on the degree of aqueous alteration of the meteorite (Takir et al. 2013). The detection of iron features in the visible range and the Mg-rich phyllosilicates band around 2750 nm point toward an heavily altered CM chondrite (Beck et al. 2010) due to a strong aqueous alteration of the meteorite on the parent body and confirms the classification of the meteorite as a CM1 or CM2 chondrite.

8. Discussion: Mukundpura compared to other CM chondrites
    a. Mukundpura as a heavily altered CM chondrite

The classification of CM chondrites into sub-groups was first proposed by McSween (1979) where the meteorites were characterized qualitatively as partially altered, altered and highly altered based on petrographical observations. Later, Browning et al. (1996) used a mineralogical alteration index describing the progressive Si and $Fe^{3+}$ substitutions forming serpentine from cronstedtite in order to quantitatively assess the level of aqueous alteration. Another scale based on several criteria was designed by Rubin et al. (2007), where petrologic subtypes from 2.0 to 2.6 were attributed to a series of CM chondrites, with the 2.0 petrologic type corresponding to the most heavily altered CM chondrites. . The Rubin scale was based on the alteration extent of chondrules (i.e. chondrule pseudomorphs), as well as the composition of tochilinite and carbonates. More recently, a classification scheme was build based to the amount of water/OH in the meteorites based on modal analysis by X-ray diffraction (Howard et al. 2015) as well bulk hydrogen measurements (Alexander et al. 2013). These scales are providing an estimate of the amount of hydrogen in the form of water of hydroxilated minerals present in the meteorite. This is somehow the extent to which the primary minerals were replaced by secondary phases, but not a measurement of water-rock ratio (which can be inferred from oxygen isotopes measurements, (Marrocchi et al. 2018)). Major advantages of the Howard et al. (2015) and Alexander et al. (2013) scales are that they are quantitative, as well as that they are applicable to CR chondrites. In that scale, petrologic types from 1.1 to 2.6 are attributed (note that the Rubin et al. (2007) and Alexander et al. (2013) are not linear).

In order to assess the level of aqueous alteration of Mukundpura, we can compare results obtained with the same techniques as the ones used in this work, on samples for which petrologic types have been described in previous works.

Table 4 compare the TGA results of Mukundpura to other CM2 chondrites.

|  | Rubin et al. 2007 scale | Alexander et al. 2013 scale | Fall/find | loss 0-200°C (wt%) | loss 200-400°C (wt%) | loss 400-760°C (wt%) | loss 760-900°C (wt%) | Total loss (wt%) |
|---|---|---|---|---|---|---|---|---|
| ALH 83100 | CM1/2.0 | 1.1 | Find | 2.5 | 2.4 | 11.0 | 0.9 | 16.2 |
| MET01070 | CM1/2.0 | 1.2 | Find | 2 | 1.4 | 9.7 | 1.2 | 14.3 |
| EET 96029 | Heated CM2 | - | Find | 5.2 | 3.6 | 4.4 | 0.7 | 15.1 |
| QUE97990 | CM2/2.6 | 1.7 | Find | 4 | 2.6 | 6.1 | 2.3 | 15 |

| WIS 91600 | Heated CM2 | - | Find | 4.9 | 0.5 | 7.2 | 2.1 | 14.3 |
| --- | --- | --- | --- | --- | --- | --- | --- | --- |
| ALH 84033 | Heated CM2 | - | Find | 4.7 | 2.9 | 3.8 | 0.9 | 13.3 |
| MAC 88100 | Heated CM2 | - | Find | 3.8 | 1.6 | 9.2 | 1.1 | 15.3 |
| MIL 07700 | Heated CM2 | - | Find | 5.4 | 1.6 | 4.3 | 0.4 | 11.4 |
| *Mukundpura* | *CM2* | - | *Fall* | *2.2* | *3.0* | *9.5* | *1.2* | *15.9* |

Table 4: TGA results of the Mukundpura meteorite compared to other CM2 chondrites (data from Garenne et al. (2014)).

The TGA results obtained on Mukundpura reveal an important mass loss (9.5 wt%) in the range associate to dehydroxilation of phyllosilicates. It seems to show the sample is dominated by phyllosilicates, which is consistent with its classification as a CM chondrite (Cloutis et al. 2011; Rubin et al. 2007). However, it is not excluded to have other contributions to this mass loss. Fe-sulphides and organics can also be decomposed in this range and contribute in the total mass loss budget. It is assumed for CM chondrites, these other contributions are low compared to the phyllosilicates abundance (Garenne et al. 2014). The mass loss is similar to the CM1/2.0 meteorite MET 01070, and so is consistent with Mukundpura being classified as a CM2.0, in (Rubin et al. 2007) scale or 1.2 CM chondrite in (Alexander et al. 2013) scale.

Figure 8 compares the mid-infrared transmission spectrum of Mukundpura to other carbonaceous chondrites of known petrologic types.

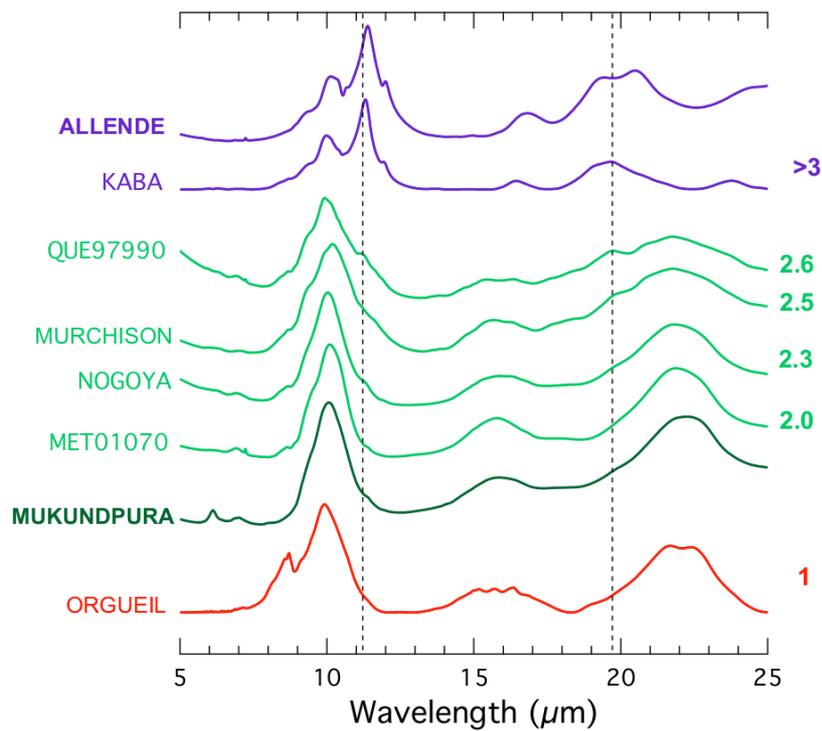

Figure 8: Infrared transmission spectroscopy of Mukundpura and other meteorites of different types, according to their group (Red: CI, Green: CM and Purple: CV) and petrologic types (right) (data from Beck et al. (2014)) The two doted lines mark the position of the absorption bands of olivine at 11.2µm and 19.5µm.

The transmission spectrum of Mukundpura is consistent with those obtained on other CM chondrites. The infrared transmission spectra of QUE 97990 (2.6), Murchison (2.5), Nogoya (2.3) and Mukundpura show faint but detectable increases of absorbance at 11.2µm and 19.5µm, due to the small amount of olivine still present in the sample after the aqueous alteration. Such features are absent in the transmission spectrum of MET 01070 (CM1)/2.0. This is indicative of a very low amount of anhydrous silicates in the sample, which is consistent with the TGA results. Figure 9 presents a comparison of the transmission spectral parameters of Mukundpura with other already classified CM chondrites. The transmission spectrum of Mukundpura showing the 2.7µm feature is displayed in appendix. The parameters of Mukundpura have been calculated using the same method as Beck et al. (2014).

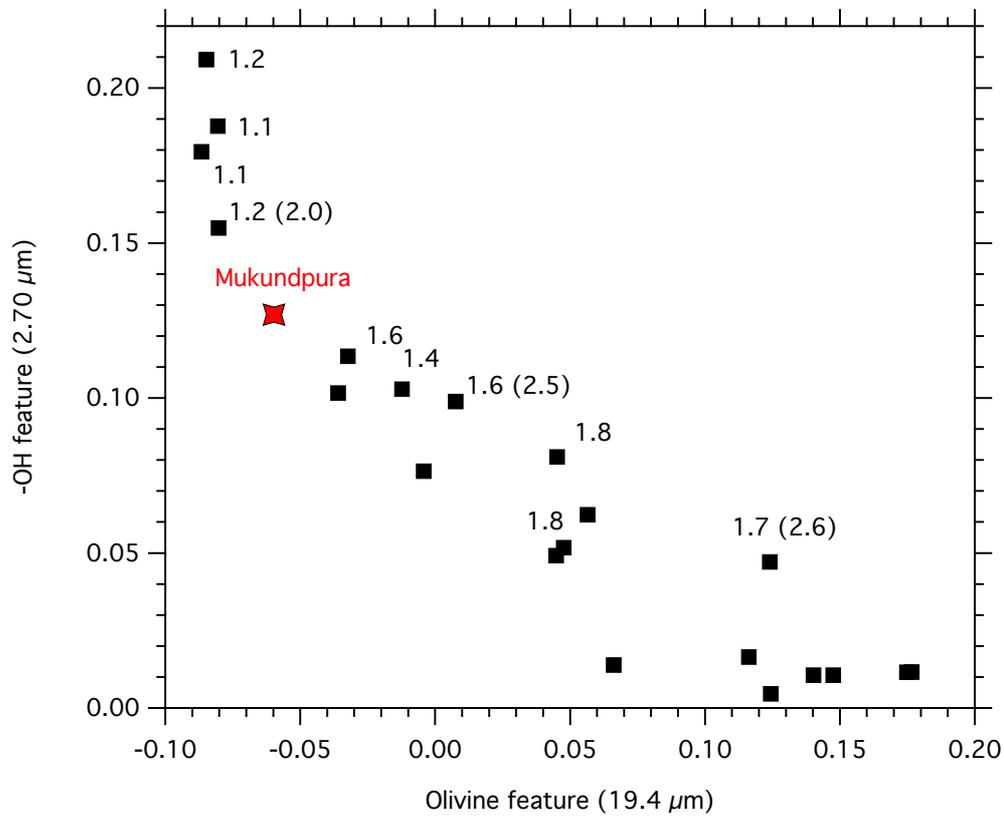

Figure 9: Correlation between the phylosilicates -OH and the olivine features in the transmission spectrum of Mukundpura, compared to other aqueously altered chondrites (data from Beck et al. (2014)). Petrologic scales displayed from Alexander et al. (2013) and Rubin et al. (2007) in parenthesis.

It shows that the meteorite is among the most aqueously altered known chondrite. Altogether TGA and transmission results reveal that Mukundpura has characteristics consistent with the more highly altered CM chondrites.

The reflectance spectroscopy of Mukundpura (Figure 7) shows all features related to CM chondrites, present in both CM1 and CM2 chondrites. The detection of spectral features alone cannot thus be used to assess the petrographic grade of a meteorite. However, as suggested in Beck et al. (2010), the shape of the 3µm band can be used as a petrographic criterion, after removing the contribution of the terrestrial absorbed water that modifies the shape of the band. We selected 9 CM chondrites of different petrographic grades to be compared to Mukundpura : ALH 83100 (CM1/1.1), ALH 84033 (CM2), EET 96029 (CM2), MAC 88100 (CM2/1.7), MET 01070 (CM1/1), MIL 07700 (CM2), Murchison (CM2/1.6), QUE 97990 (CM2/1.7) and WIS 91600 (CM2). The reflectance spectra were acquired under secondary vacuum to evacuate the adsorbed water, as described in Potin et al. (under review), and are presented on Figure 10.

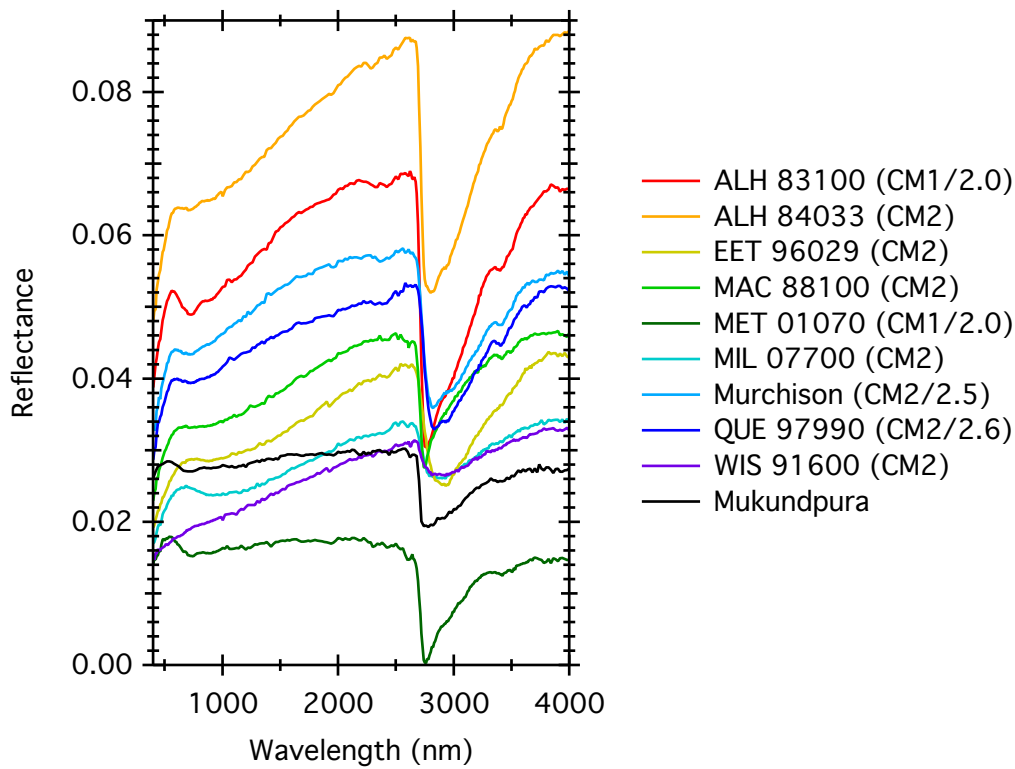

Figure 10: Reflectance spectra of the comparative CM chondrites considered for comparison with Mukundpura (data from Potin et al. (under review)). For clarity, a vertical offset of -0.02 and +0.02 has been applied respectively to MET 01070 and ALH 84033.

Figure 11 compares the shape of the phyllosilicate band of Mukundpura with the other CM chondrites. The barycenter of the band was calculated between 2600nm and 3000nm as:

$$B_{\lambda=2600}^{\lambda=3000} = \frac{\sum_{\lambda=2600}^{\lambda=3000}(R_\lambda - C_\lambda)\lambda}{\sum_{\lambda=2600}^{\lambda=3000}(R_\lambda - C_\lambda)}$$

with $R_\lambda$ and $C_\lambda$ respectively the measured reflectance and the calculated continuum at the wavelength λ (Beck et al. 2010; Pommerol et al. 2009). The Full Width Half Maximum of the band is considered as the horizontal width of the band, measured at the midpoint between the continuum and the minimum of reflectance (Cloutis et al. 2015).

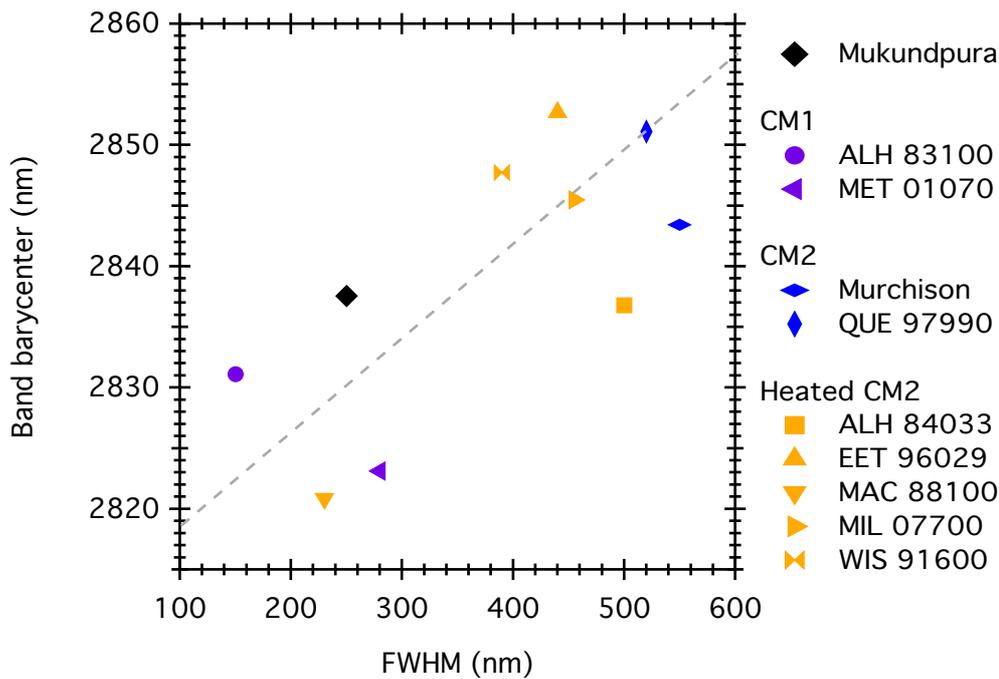

Figure 11: Barycenter and full width at half maximum (FWHM) of the 3μm band of Mukundpura and other aqueously altered chondrites. Error bars are smaller than the size of the markers.

The increasing degree of alteration affects the reflectance spectra by thinning and shifting the phyllosilicates band towards shorter wavelengths. The 3µm band of the reflectance spectrum of Mukundpura presents a FWHM of 250 nm and a barycenter at 2837 nm, meaning that the shape of the band is similar to the one observed in other spectra from CM1/2.0 chondrites ALH 83100 (FWHM of 150nm, barycenter at 2831nm) and MET 01070 (FWHM of 280nm and barycenter at 2823nm). Using the shape of the phyllosilicates band as a criterion, it is justified to classify Mukundpura as a CM1/2.0 chondrite.

b. Something special with the reflectance spectroscopy of a fresh CM chondrite?

Mukundpura was rather well protected from post-fall terrestrial alteration so by using this meteorite, the reflectance spectroscopy of a fresh chondrite can therefore be compared to other known samples. We used as comparison parameters the value of reflectance (measured at 1500 nm outside of an absorption band), the spectral slope described as the ratio between the reflectance measured at 2200 nm and at 1100 nm, and the depth of the visible, 3µm and organic absorption features. The depth of an absorption band is calculated assuming a linear continuum between the two inflection points around the feature, and using the following equation

$$BD = 1 - \frac{R^{Band}{}_\lambda}{R^{Continuum}{}_\lambda}$$

with $R^{Band}{}_\lambda$ and $R^{Continuum}{}_\lambda$ being respectively the measured reflectance at the center of the band (at the wavelength $\lambda$) and the calculated reflectance of the continuum. The comparisons are presented on Figure 12.

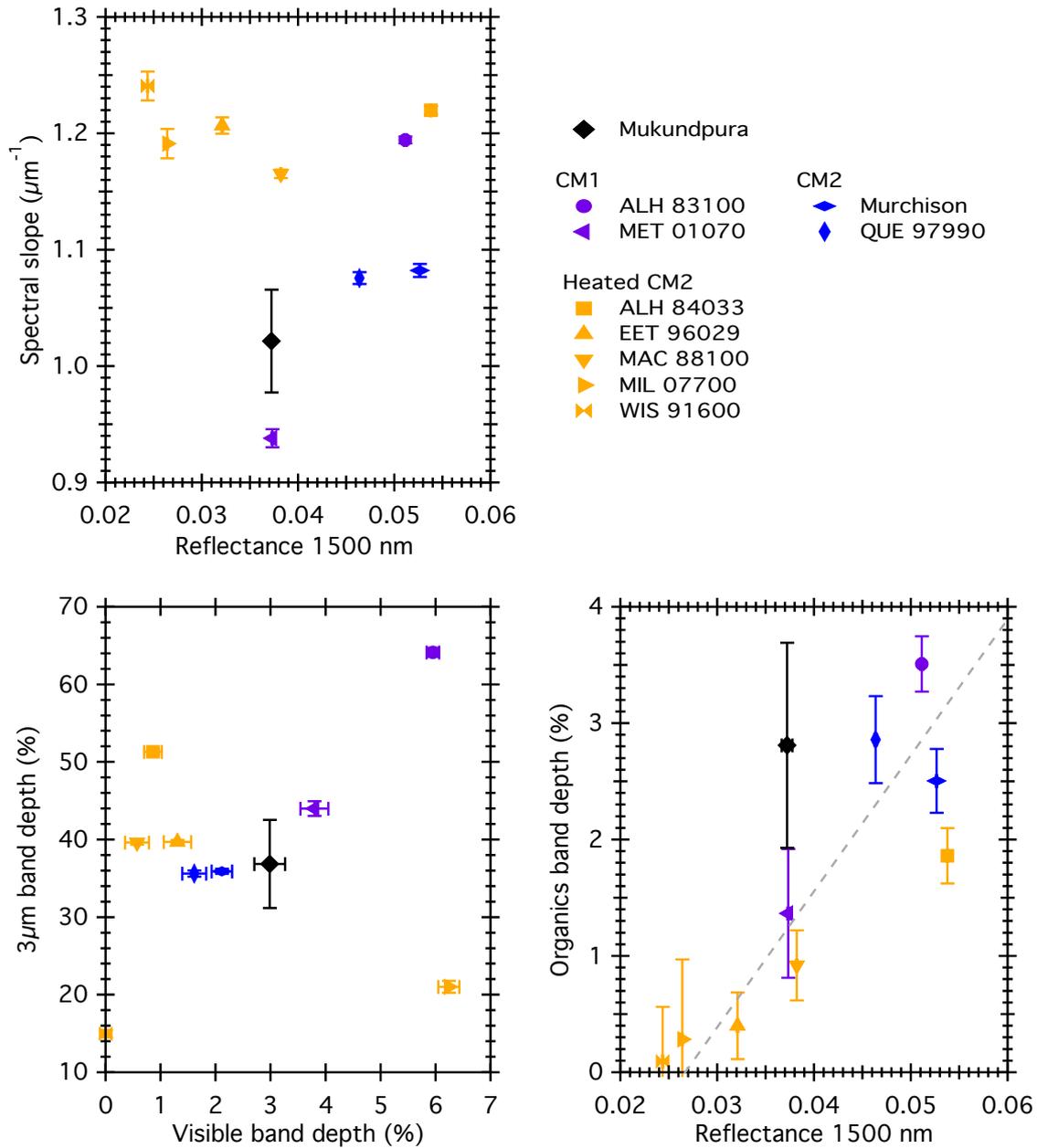

Figure 12: Comparison of the spectral parameters of Mukundpura with other CM chondrites.

Mukundpura shows a reflectance value similar to other CM chondrites. The meteorite presents a faint spectral slope of 1.02 ± 0.04μm$^{-1}$, lower than the other samples that thus appear redder. Only MET 01070, one of the most altered chondrites in this analysis shows a weaker spectral slope. The bands depths of Mukundpura match the other CM1 and unheated CM2 chondrites, at least for the visible 700nm and the 3μm bands.

Figure 12 shows positive correlation between the reflectance value and the depth of the organic band. This is consistent with the decrease of the absorption band depth along with the increasing amount of opaque minerals and dark material, whose presence in the sample tends to lower the reflectance of the surface (Pommerol and Schmitt 2008). Mukundpura seems to be excluded from the general trend with a deeper organic feature compared to the other meteorites with similar measured reflectance. As some of the organic matter in carbonaceous chondrites is the most likely material to be destroyed by weathering conditions (Bland et al. 2006), the reflectance spectra of falls should present deeper organic features than finds, as is the case for the freshly fallen Mukundpura. Note however, that the trend in Figure 12 relies upon CM and heated CM, and more measurements on non-heated CM are required to confirm the trend. While Mukundpura share mineralogical affinities with CM1, previous studies of CM1 organics have revealed that they are depleted in organics compared to CM2 (Aponte et al. 2011; Glavin et al. 2012) as well as depleted in C content (Alexander et al. 2013). However, the C content measured for Mukundpura (2.3 wt. %) is typical (if not high) for CM2 chondrites (Rudraswami et al. 2019).

9. Summary and conclusion

We conducted several analyses to investigate the alteration history of the Mukundpura meteorite. The results are summarized below.

- Infrared transmission spectroscopy shows a faint olivine signature and indicates, along with the thermogravimetry analysis, that Mukundpura is dominantly

- composed of phyllosilicates. This indicates a strong alteration of the minerals on the parent body.
- Raman spectroscopy shows a low degree of graphitization of the polyaromatic carbonaceous matter in the matrix, which reflects the absence of significant thermal or shock metamorphism.
- Reflectance spectroscopy of Mukundpura shows all spectral features of an aqueously altered CM chondrite, such as the $Fe^{2+}$ and $Fe^{3+}$ absorption bands at 700nm, 900nm and 1100nm, the –OH stretching band of phyllosilicates around 2750nm and a faint organic feature at 3400nm.
- Mukundpura is a fresh fall and thus differs from the other studied chondrites. The reflectance spectrum of Mukundpura presents one of the lowest spectral slopes. While the reflectance is lower that other CM chondrites studied here, the organic band depth is similar, which may suggest that the sample is unusually rich in organics.

The classification of Mukundpura as an aqueously altered chondrite is well established. In this work we have investigated a single piece of the Mukunndpura meteorite fall, and since CM meteorite are breccia the conclusion reached here may not apply to the meteorite as a whole. Our TGA, infrared spectroscopy and reflectance spectroscopy are consistent with Mukundpura being a CM1/2.0, rather than a CM2 chondrite.

Aknowledgments

SP is supported by Université Grenoble Alpes (IRS/IDEX). PB acknowledges funding from the H2020 European Research Council (ERC) (SOLARYS ERC-CoG2017_771691). AG's participation in the preparation of this manuscript was performed under the auspices of the U.S. Department of Energy by Lawrence Livermore National Laboratory under Contract DE-AC52-07NA27344, release number LLNL-JRNL- 973015. AKM thanks the AvH foundation for support during his stay in Germany. The instrument SHADOWS was founded by the OSUG@2020 Labex (Grant ANR10 LABX56), by 'Europlanet 2020 RI' within the European Union's Horizon 202 research and innovation program (grant N° 654208) and by the Centre National d'Etudes Spatiales (CNES).

APPENDIX A:

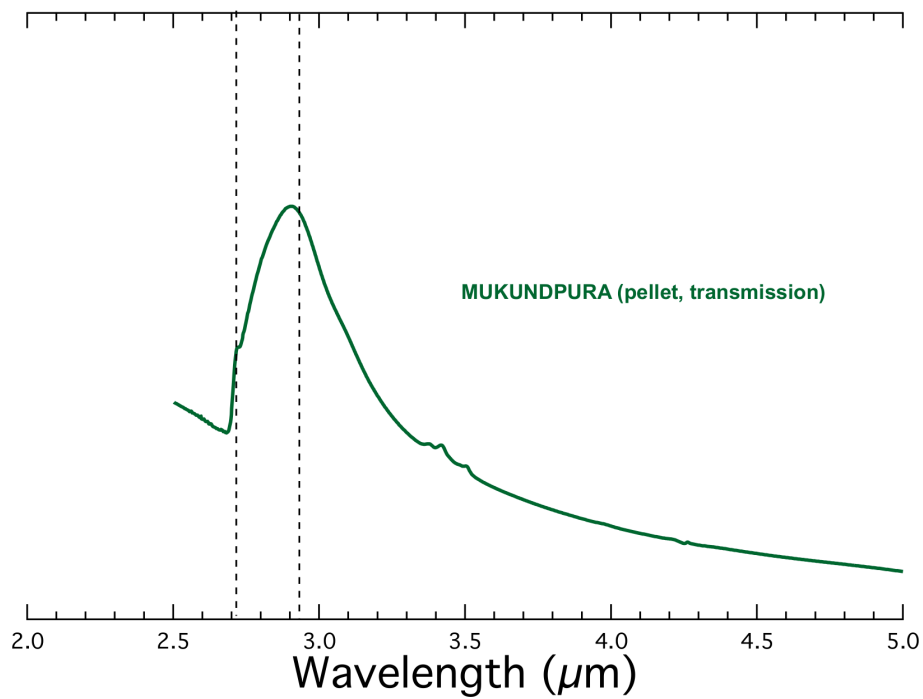

Figure 13 : KBr pellet measurement of a small fraction the Mukundpura meteorite in the 3-μm region. The presence of OH is indicated by the presence of a narrow feature around 2.7 μm micron. This narrow feature is on top of a broader feature with a maximum at 2.9 μm

chondritic meteorites. *Geochimica et Cosmochimica Acta* 31:747–765.

http://linkinghub.elsevier.com/retrieve/pii/S0016703767800309.